\documentclass[twocolumn,showpacs,preprintnumbers,amsmath,amssymb]{revtex4}

\input epsf

\usepackage{graphicx}% Include figure files
\usepackage{dcolumn}% Align table columns on decimal point
\usepackage{bm}% bold math

\begin{document}

%\preprint{}

\title{Phase transitions on Markovian bipartite graphs---an application of 
the zero-range process}

\author{Otto Pulkkinen}
\email{otto.pulkkinen@phys.jyu.fi}
\author{Juha Merikoski}
\email{juha.merikoski@phys.jyu.fi} 
        
\affiliation{
Department of Physics, P.O.~Box 35 (YFL), FIN-40014 
University of Jyv\"askyl\"a, Finland}

\date{\today}

\begin{abstract}

We analyze the existence and the size of the giant component in the stationary 
state of a Markovian model for bipartite multigraphs, in which the movement 
of the edge ends on one set of vertices of the bipartite graph is a 
zero-range process, the degrees being static on the other set. 
The analysis is based on approximations by independent variables 
and on the results of Molloy and Reed for graphs with prescribed degree 
sequences. The possible types of phase diagrams are identified by studying 
the behavior below the zero-range condensation point. As a specific example, 
we consider the so-called Evans interaction. In particular, we examine 
the values of a critical exponent, describing the growth of the giant 
component as the value of the dilution parameter controlling the 
connectivity is increased above the critical threshold. Rigorous analysis 
spans a large portion of the parameter space of the model exactly at the 
point of zero-range condensation. These results, supplemented with 
conjectures supported by Monte Carlo simulations, suggest that the 
phenomenological Landau theory for percolation on graphs is not broken by the 
fluctuations.

\end{abstract}


%%% 02.50.Ey Stochastic processes
%%% 05.40.-a Fluctuation phenomena, random processes

\maketitle

%%%%%%%%%%%%% INTRO

\section{\label{intro}Introduction}

The common way of implementing time dependence in the 
models of random graphs is to couple the time and the 
number of edges in such a way that, starting with an empty graph with a 
given number of vertices, a new edge is added uniformly at random 
to one of the vacant edge locations at each time step. This procedure is 
known simply as the {\it graph process} \cite{Bollobas85}. During the past few 
years, a large variety of models generalizing the graph process have appeared 
in the physics literature (see Ref.~\cite{Dorogovtsev02} for a review), most 
of them motivated by the structure of the Internet and the World Wide Web.
One of the central ideas of these models is known, by the work of 
Bar\'abasi and Albert \cite{Barabasi99}, as preferential 
attachment. In the Barab\'asi-Albert model new vertices are introduced one at 
a time and joined to one of the existing vertices chosen with probability 
proportional to the {\it degree} of the vertex, i.e.\ the number of edges 
already joined to it. This preferential attachment leads to what 
Price \cite{Price76} calls cumulative advantage, and it can be used to 
construct so-called scale-free networks, i.e. graphs with power-law 
distributed degrees.   

In the graph process and the Barab\'asi-Albert model, the edges, once 
attached to the vertices, retain their positions {\it ad infinitum}. 
In the present article, we consider what 
happens if internal restructuring by rewiring of the graph with the existing 
edges dominates the dynamics, that is, the time scales associated with 
addition and removal of edges and vertices, have become very long as compared 
to the times between the rewiring events. The dynamics is taken to be the 
simplest possible capable of providing the cumulative advantage - 
namely of zero-range \cite{Spitzer70} so that the rate of a rewiring 
event, i.e.\ a jump of an edge end from one vertex to another, 
depends only on the degree of either the initial or the final 
vertex of the move. We focus on the connectivity properties of the graphs in 
the stationary state. The question 
about the existence and size of the giant component, a connected cluster 
occupying a macroscopic fraction of vertices, turns out particularly 
interesting because of the condensation phenomena induced by the 
zero-range processes. Closely related models have been studied on a 
phenomenological level by Palla {\it et al.}~\cite{Palla04} and rigorous 
results for models of polymerization sharing the same kind of ideas 
have been obtained by Pittel {\it et al.}~\cite{Pittel90a,Pittel90b}.
A statistical mechanics approach to a rather general class of reversible 
graph-valued processes is presented in Ref.~\cite{Dorogovtsev03}.

We take bipartite multigraphs as the state space of 
the problem so that the system consists of external agents with static 
degrees and another set of vertices, for example collaboration events, on 
which the movement of the edge ends, coming from the set of agents, takes 
place. A process with unipartite states is obtained as a special case to 
our model, but its behavior is not as rich as in the bipartite version. 

The structure of the article is as follows: In the following section, 
we give a precise definition of the model and discuss the reasons for 
choosing the particular state space. Section \ref{SecZRP} reviews some 
fundamental results about zero-range processes and a Poisson approximation 
for the numbers of vertices of given degree is constructed. This is applied to 
subcritical zero-range processes in section \ref{SecSub}. In section 
\ref{SecExamples}, two simple examples are given and general implications 
of the results for subcritical zero-range processes are discussed in section 
\ref{SecGen}. In section \ref{SecEvans}, we concentrate on restructuring 
processes with the so-called Evans interaction. A detailed view of the phase 
diagram and the critical exponents is given in subsection \ref{SecPhase}. 
The connection with a model of diluted scale-free networks is also 
established. Conclusions are made and some open problems presented in 
section \ref{SecConclusions}.

\section{Statement of The Problem}
\protect\label{SecProblem}

\begin{figure}
\includegraphics[clip,angle=0,width=0.5\textwidth]{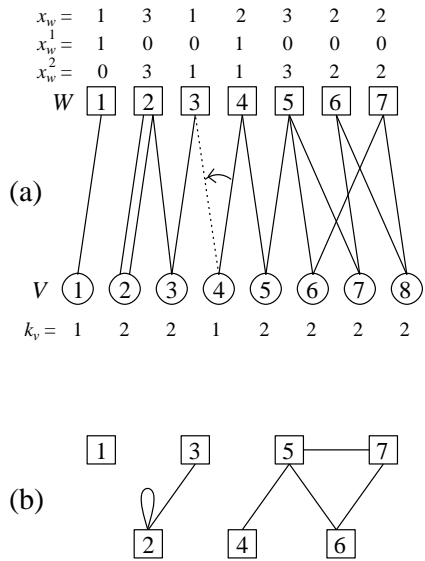}
\caption{\label{fig1}(a) Example of a bipartite multigraph with vertex sets 
$V$ and $W$. The static degrees of $v\in V$ are denoted by $K_v$ and the 
evolving degrees of $w\in W$ by $X_w = X_w^1 +X_w^2$ with $X_w^1$ and 
$X_w^2$ the numbers of edges leading to vertices of degree one and two, 
respectively. The arrow and the dotted line show a possible transition.   
(b) The projection of the bipartite graph in (a) to the vertex set $W$.}
\end{figure}

We shall consider bipartite multigraphs with vertex sets $W$ and $V$ such that 
$|W| = L$ and $|V| = M_1 + M_2$ (see Fig.~\ref{fig1}(a)). 
Here $M_1$ denotes the number of vertices of 
degree $1$ and $M_2$ the number of vertices of degree $2$ and these degrees 
are assigned to randomly chosen vertices in $V$ independently of each other. 
This set up on $V$  will be static during the time evolution. 
The total number of edges is $N = M_1 + 2M_2$ and we shall consider only cases 
with the densities 
\begin{equation}
\label{rrho}
r:= 2M_2 /N\in \lbrack 0,1 \rbrack, \qquad \rho := N/L \in (0,\infty )
\end{equation}
fixed on the passage to the limit $L \to \infty$. 

Let us then discuss the dynamics.
Initially, the $N$ edge ends are chosen to be distributed uniformly and 
independently of each other on the vertices in $W$. The restructuring of the 
graph takes place in continuous time by jumps of the edge ends on this set. 
We take the process on $W$ to be a zero-range process 
\cite{Spitzer70,Evans00} with $M_1$ particles of type $1$ (edges to vertices 
of degree $1$ in $V$) and $2M_2$ particles of type $2$ (edges to vertices 
of degree $2$ in $V$), so that a vertex $w\in W$ loses a randomly 
chosen edge end after an exponentially distributed time, where the parameter 
of the exponential distribution is a function of the degree of $w$. More 
precisely, if $w\in W$ is an endvertex of $X_w^1$ edges to vertices of 
degree $1$ and of $X_w^2$ edges to vertices of degree $2$, the 
exponential rate is given by $g(X_w^1 + X_w^2)$, where $g$ 
is a given positive function with bounded increments, and the edge end to 
jump is one of those leading to vertices of degree $1$ with probability 
$X_w^1/(X_w^1 + X_w^2)$ and one of those leading to vertices of degree $2$ 
otherwise. Immediately after leaving the vertex $w$, the edge end 
joins one of the other vertices according to a symmetric, irreducible 
transition matrix $(P_{w,z})$, $w,z\in W$. The stationary distribution of 
this zero-range process can be found explicitly and will be studied in the 
next section. Clearly, the stationary distribution of the corresponding 
bipartite multigraph -valued process is such that all the states with the 
same degree sequence are equiprobable. 

In this article, we study the structure of the graphs in the stationary state 
of the process. In particular, we concentrate on the properties of the 
projection of the bipartite graph, also called one-mode network 
\cite{Newman01}, on the set $W$, obtained by adding 
an edge between two vertices in $W$ if one can be reached from the other by 
traversing two edges on the original bipartite graph, see Fig.~\ref{fig1}(b). 
In other words, the projection tells us how the vertices in $W$ are mutually 
connected by the vertices of the other type. The particular questions we ask 
are then, when does a giant component, a connected subgraph of size 
proportional to $L$, exist on the projection and precisely of how many 
vertices such a component is made? 

There are several reasons for making the particular choices concerning the 
state space. First of all, the zero-range dynamics is easily implemented in 
this setting and it contains the unipartite model as the special case $M_1=0$.
The bipartite structure really brings something more to the theory: The $M_1$ 
vertices of degree $1$ are dangling ends from the point of view of 
connectivity, and do not appear on the projection graph, but they do have 
influence on the jump rates of the zero-range process. The parameter 
$r=2M_2/N$ controls the amount of this 'dark matter' and lowering its value 
while keeping $\rho = N/L$ fixed corresponds to diluting the set of edges 
on the projection. Therefore it will be referred to as a dilution parameter. 
Varying the values of $r$ and $\rho$ simultaneously 
in such a way that their product remains constant can be used to sweep the 
phase diagram of the underlying particle system, characterized by the density 
$\rho$ alone, without changing the density of edges on the projection. 
In addition, the restriction of the static degrees 
to at most $2$ allows us to use the usual configuration 
model \cite{Bollobas85}, and therefore the results of Molloy and Reed 
\cite{Molloy95,Molloy98}, directly without going through the proofs in the 
bipartite setting. Moreover, we believe that the extension that allows higher 
but bounded degrees or even degree distributions with exponential tails 
does not change the qualitative features of the phase diagrams or the values 
of the critical exponents.     

\section{Zero-range processes and Poisson approximations}
\label{SecZRP}

Let the dynamics of the edge ends on the set $W$ be as described in the 
previous section and let $\Omega_{L,M_1,M_2}$ be the set of ordered 
partitions of two numbers $M_1$ and $2M_2$ both in $L$ non-negative parts. 
Then, for 
$(\eta ,\xi )\in \Omega_{L,M_1,M_2}$, one shows by verifying that the detailed 
balance condition holds that the stationary (canonical) distribution of the 
zero-range process is given by
\begin{eqnarray}
\label{mu}
& & \mu_{L,M_1,M_2} (\eta ,\xi ) = \\ \nonumber
& & \ \ \ \frac{1}{Z_{\mu} (L,M_1,M_2)} \prod_{w=1}^L 
\binom{\eta_w +\xi_w}{\eta_w} \frac{1}{g!(\eta_w +\xi_w)},
\end{eqnarray}
where $g!(n) = \prod_{k=1}^n g(k)$ is a generalized factorial with the 
convention that $g!(0) = 1$ and $Z_{\mu} (L,M_1,M_2)$ is a normalization 
factor (canonical partition function). 
Let $D_{m,n}$ denote the number of vertices in 
$W$ with exactly $m$ edges leading to vertices of degree $1$ and $n$ edges 
leading to vertices of degree $2$. Then (\ref{mu}) implies an image measure 
\begin{eqnarray}
\label{sigma}
& & \sigma_{L,M_1,M_2} (d) = \frac{1}{Z_{\sigma} (L,M_1,M_2)} 
\\ \nonumber
\ \ \ \  & & \times \prod_{m=0}^{M_1}
\prod_{n=0}^{M_2} \frac{1}{d_{m,n}!}\left( \binom{m + n}{m} 
\frac{1}{g!(m + n)} \right)^{d_{m,n}}
\end{eqnarray}
for the joint random variable $D = (D_{m,n})$ in the stationary regime. 
In Eq.~(\ref{sigma}), $d$ is such that the three conditions 
$\sum_{m=0}^{M_1} \sum_{n=0}^{2M_2} d_{m,n} = L$, 
$\sum_{m=0}^{M_1} \sum_{n=0}^{2M_2} (m+n) d_{m,n} = M_1+2M_2$ and
$\sum_{m=0}^{M_1} \sum_{n=0}^{2M_2} m d_{m,n} = M_1$ are satisfied. 
Otherwise $\sigma_{L,M_1,M_2}$ vanishes. 

Our purpose is to get a grip of the random variables 
$D_n := \sum_{m=0}^{M_1} D_{m,n}$, i.e.\ the numbers of vertices of degree 
$n$ on the projection to $W$. More formally, 
${\mathcal D} = (D_n)$ is the asymptotic degree sequence of Molloy and Reed 
\cite{Molloy95,Molloy98} in our problem and we have to show that, with high 
probability \cite{whp}, we get an asymptotic degree sequence 
${\mathcal D}$, which is in their sense well-behaved (see appendix \ref{AppA} 
for definitions and discussion). Because of the constraints on $d$ in 
Eq.~(\ref{sigma}), the random variables $D_{m,n}$ 
are not independent, and it would be difficult to extract the behavior of 
$D_n$ from the measure $\sigma_{L,M_1,M_2}$ directly. Next we are going to 
construct a sequence of independent variables that is shown to be a good 
approximation for the original sequence. In other words, we shall study the 
problem in the grand canonical ensemble and prove that the canonical and 
grand canonical ensembles are equivalent.   

For zero-range processes, approximations by independent variables have been 
considered by Gro{\ss}kinsky {\it et al}.~\cite{Grosskinsky03} and 
Jeon {\it et al}.~\cite{Jeon00}. In their highly influential article, 
Jeon {\it et al}.~were able to give a rigorous proof for the behavior of the 
largest cluster not only for subcritical processes and for the cases with 
$b>3$ in what is sometimes called Evans interaction \cite{Evans00}, 
$g(k) = 1 +b/k + O(k^{-(1+\delta)})$, but also for a large 
class of vanishing rates. Within this class, which essentially consists of 
functions that vanish faster than $\exp(-c\sqrt{\log k})$, a single vertex 
would hold all but of the order $Lg(L)$ edge ends in our model, so that 
the projection of the graph would be in a flower-like state with one vertex 
having many self-loops. We choose to study the vanishing cases no further. 
By restricting ourselves to the cases 
with the interaction function $g$ bounded away from zero, we also avoid the 
truncation of the series involved in the grand canonical analysis. 
The phase diagram of the Evans model was also discussed by 
Gro{\ss}kinsky {\it et al}.~and heuristic arguments, supported by simulation 
results, for the dynamics of condensation were given. 
The results of these articles concerning the Evans 
interaction will be applied in later sections. 
The general method and the validity of approximations by independent variables 
is discussed in a delightful paper of Arratia and Tavar\'e \cite{Arratia94}.

Let now $g$ be bounded away from zero and let $C_{m,n}$, $m,n\in 
{\mathbb Z}_+$, 
be independent 
$\mathrm{Poisson}(\lambda_{m,n}(\alpha,\gamma,\phi))$-distributed 
random variables, where
\begin{equation}
\label{lambda}
\lambda_{m,n}(\alpha,\gamma,\phi) = \frac{1}{Z(\phi)} \binom{m+n}{m} 
\frac{ \alpha (1-\gamma)^m \gamma^n \phi^{m+n}}{g!(m+n)}, 
\end{equation}
\begin{equation}
Z(\phi) = \sum_{k=0}^{\infty} \frac{\phi^k}{g!(k)},
\end{equation}
and $\alpha,\gamma$ and $\phi$ are real parameters whose values will be 
determined later. Here $Z$ is the grand canonical partition function. Its 
radius of convergence, which we shall denote by $\Phi$, is positive because 
$g$ is bounded away from zero. 
In order to avoid cumbersome expressions, we set 
\begin{eqnarray}
\label{conditions}
A_1 &=& \Big\lbrace \sum_{m=0}^{\infty} \sum_{n=0}^{\infty} C_{m,n} = L 
\Big\rbrace, \nonumber \\
A_2 &=&  \Big\lbrace \sum_{m=0}^{\infty} \sum_{n=0}^{\infty} (m+n) C_{m,n} = 
N \Big\rbrace, \\ 
A_3 &=&  \Big\lbrace \sum_{m=0}^{\infty} \sum_{n=0}^{\infty} m C_{m,n} = 
M_1 \Big\rbrace. \nonumber
\end{eqnarray}
Then one can show by a direct calculation that 
$( D_{0,0}, \ldots ,D_{M_1,2M_2}) 
= ( C_{0,0} , \ldots \vert A_1\cap A_2\cap A_3)$ in law and, because of 
cancellation in the conditional probabilities, independent of the values of 
$\alpha,\gamma$ and $\phi$. So $(C_{m,n})$ really is a candidate for a good 
approximation. The construction suggests the following for the approximative 
system: 
\begin{enumerate}
\item{A direct calculation shows that the system size
$\sum_{m=0}^{\infty} \sum_{n=0}^{\infty} C_{m,n}$ 
is a $\mathrm{Poisson}(\alpha)$-distributed random variable.}
\item{Again by a direct computation, for $t\in {\mathbb R}$, 
\begin{eqnarray*}
& & E \exp \left( it\sum_{m=0}^{\infty} \sum_{n=0}^{\infty} (m+n)
  C_{m,n} \right) = 
\\ \nonumber
& & \ \ \ \ E\left( \frac{Z(\phi e^{it})}{Z(\phi )}\right)^
{\sum_{m=0}^{\infty} \sum_{n=0}^{\infty} C_{m,n}},
\end{eqnarray*}
in that 
\begin{equation*}
\left(\sum_{m=0}^{\infty} \sum_{n=0}^{\infty} (m+n) C_{m,n} 
\bigg\vert A_1 \right) = \sum_{w=1}^L Y_w 
\end{equation*}
in law, with the variables $Y_w$ independent and distributed according to 
the grand canonical distribution
\begin{equation}
\nu_{\phi}(k) :=  \frac{1}{Z(\phi)} \frac{ \phi^{k}}{g!(k)}, \qquad 
k\in{\mathbb Z}_+.
\end{equation}}
In other words, $\nu_{\phi}$ controls the vertex degrees locally. 
The expectation of a $\nu_{\phi}$-distributed variable will be 
denoted by $R(\phi)$:
\begin{equation}
\label{defphi}
R(\phi) := E Y_w = \phi Z'(\phi)/Z(\phi).
\end{equation}
The parameter $\phi$ is known as the fugacity. 
\item{In a similar manner one shows that 
\begin{eqnarray*}
E \bigg\lbrack \exp \bigg( it\sum_{m=0}^{\infty} \sum_{n=0}^{\infty} 
m C_{m,n}\bigg) \bigg \vert A_1 \bigg\rbrack = \\
  E\left( \gamma + (1-\gamma)e^{it} \right)^{\sum_{w=1}^L Y_w}, 
\end{eqnarray*}
so that, given $A_1$ and $A_2$, the number of edges leading to vertices 
of degree equal to unity has the binomial distribution with parameters $N$ 
and $\gamma$}.  
\end{enumerate} 

Now suppose that one would like to get an exponentially decaying upper bound 
for the probability of an event $B$ concerning the dependent variables 
$(D_{m,n})$, and let ${\mathcal B}$ be the same event for the independent 
variables $(C_{m,n})$. Suppose further that the parameters $\alpha,\gamma$ 
and $\phi$ can be chosen in such a way that the probability of the 
conditioning event $A_1\cap A_2 \cap A_3$ is not exponentially small. 
Then it suffices to get the upper bound for the independent case:
$P(B) = P( {\mathcal B}\vert A_1\cap  A_2\cap  A_3) \leq 
P({\mathcal B})/P(A_1\cap  A_2\cap  A_3)$. The use of this conditioning 
device is nicely illustrated in the articles of Corteel {\it et 
al}.~\cite{Corteel99} and Jeon {\it et al}.~\cite{Jeon00}. 

How to best choose the values of the parameters? The point $1$ in the list 
above tells that the approximative system size 
$\sum_{m=0}^{\infty} \sum_{n=0}^{\infty} C_{m,n}$ is sharply peaked 
at $\alpha$ with a standard deviation of $\sqrt{\alpha}$ so that the simplest 
choice is to match the parameter with the size of the canonical system $L$. 
Notice that the choice is not unique because terms of the order $o(L)$ 
could be added without affecting the behavior in the large system size limit.  
The points $2$ and  $3$ in the list show that, with the appropriate 
conditionings, the same type of argument applies to the two other random 
variables in the conditions (\ref{conditions}) as well. We thus set
\begin{equation}
\label{setalphar}
\alpha = L, \qquad \gamma = 2M_2/N = r 
\end{equation}
and, if possible, 
\begin{equation}
\label{setphi}
R(\phi) = N/L = \rho.
\end{equation} 
Clearly, the first two equations can 
always be fulfilled but in the third one the case 
\begin{equation}
\label{rhoc}
\rho_c := \lim_{\phi\to \Phi} R(\phi) < \rho 
\end{equation}
marks a condensation transition. Roughly speaking, the best one can do in 
that case is to set $\phi =\Phi$, so that locally one observes degrees 
distributed according to $\nu_\Phi$, just like in a system with density 
$\rho_c$. Then the rest of the mass, that is $(\rho - \rho_c)L$ 
edges, must hide in the $o(L)$-sets of vertices since those edge ends are not 
picked up by the approximating measure. It was proven in 
Refs.~\cite{Jeon00,Grosskinsky03} that this really is the true picture and, 
furthermore, the authors of Ref.~\cite{Grosskinsky03} claim that the set of 
$o(L)$ vertices with the rest of the mass is in fact a single 
vertex for a large class of rate functions. Especially, the Evans model, 
which we shall study in section \ref{SecEvans}, belongs to this class. 
As a by-product of the discussion above, we get something really important 
to the theory. From the point of view of our application the condensates for 
vanishing rates and the ones bounded away from zero are completely different: 
In the first case, 
the components on the flower-like  projection graph are very small, while 
in the second, the condensate on a single vertex implies the existence of a 
giant component. 

Recall that $D_n := \sum_{m=0}^{M_1} D_{m,n}$ is the number of vertices of 
degree $n$ on the projection, so that 
$C_n := \sum_{m=0}^{\infty} C_{m,n}$ is an independent approximation for this 
quantity. We define 
\begin{equation}
\label{nur}
\nu_{\phi}^r (n) =  \frac{1}{Z(\phi)} \sum_{k=n}^{\infty} \binom{k}{n} r^n 
(1-r)^{k-n}\frac{ \phi^{k}}{g!(k)},
\end{equation}
which equals the expected value of $C_n$ divided by the system size. 
To show that the conditions of Molloy and Reed are satisfied, we first prove 
that the fluctuations $C_n$ around $\nu_{\phi}^r (n) L$ are not too large.   
The following inequality is rather easily obtained for any $\epsilon > 0$ 
using Chernoff's bounds (see e.g.~\cite{Bremaud99}):
\begin{equation}
\label{Chernoffbound}
P\left( \left\vert \frac{C_n}{L} - \nu_{\phi}^r (n) \right\vert > 
\epsilon \right) \leq \exp\left( -\epsilon L f(\epsilon /\nu_{\phi}^r (n))
\right),
\end{equation}
where $f(x) =\left( (1+x)\log(1+x) -x \right)/x$, with $x>0$, is strictly 
increasing. Using here the trivial bound 
$f(\epsilon /\nu_{\phi}^r (n)) \geq f(\epsilon )$ and the fact that 
$\epsilon f(\epsilon ) \geq  (1+\epsilon) \epsilon/(1+\epsilon/2) 
-\epsilon = \epsilon^2/(2+\epsilon)$ yields that, with high probability 
for any $\delta > 0$, 
\begin{equation}
\label{Cnfluct}
C_n = \nu_{\phi}^r (n) L + o(L^{1/2 +\delta}).
\end{equation}
In order to get the estimates for the original sequence, one has to bound 
$P( A_1\cap  A_2\cap  A_3)$ from below.

\section{Subcritical zero-range processes} 
\label{SecSub}

Let us now derive a subexponential lower bound for the probability of the 
conditioning event in the subcritical cases $\rho_c := 
\lim_{\phi\to \Phi} \phi Z'(\phi)/Z(\phi) > \rho$. Notice first that with 
the choices made in Eqs.~(\ref{setalphar}) and using the results for the 
variables in $A_1$ and $A_3$, the probability of $A_1$ is bounded below by 
$c_1/\sqrt{L}$ for some constant $c_1$ and also $P(A_3 \vert A_1\cap A_2) 
\geq c_3/\sqrt{L}$. So we have to bound the probability of $A_2$ given $A_1$. 
By subcriticality and from the fact that $\phi Z'(\phi)/Z(\phi)$ is strictly 
increasing it follows that there is a unique $\phi<\Phi$ for every 
value of the density such that Eq.~(\ref{setphi}) is satisfied and the 
distribution $\nu_{\phi}$ therefore has an exponential tail. 
Therefore the usual local limit theorem \cite{Gnedenko68} applies and the 
probability is at least $c_2/\sqrt{L}$. 
Thus $P( A_1\cap  A_2\cap  A_3) \geq cL^{-3/2}$ and the bound (\ref{Cnfluct}) 
for the fluctuations of $C_n$ applies to $D_n$ as well. 
Moreover, the tail of $\nu_{\phi}^r$ is 
exponentially thin  because  $\nu_{\phi}^r (n) = O(  \nu_{\phi} (n)/r)$ and 
$\nu_{\phi}$ shows exponential decay. Also, since $\sum_{k\geq n} 
\nu_{\phi} (k) = O(c^n)$, for some $c\in (0,1)$ and large $n$, the largest 
degree is $O(\log L)$ with high probability. Now it is easy to check that, 
with high probability, the degree sequence $(D_n)$ is such that the 
conditions of Molloy and Reed, reproduced in appendix \ref{AppA}, are 
satisfied.

Next we give the results for the subcritical cases. 
Let $\Phi >0$ and $\rho_c >\rho$ and define
\begin{equation}
\label{rc}
r_c(\phi) := \frac{Z'(\phi)}{\phi Z''(\phi)}.
\end{equation}
Then, the parts of Theorem $1$ of Ref.~\cite{Molloy95} concerning the 
existence of the giant component and Theorem $1$ of Ref.~\cite{Molloy98} 
state that, with high probability:
\begin{enumerate}
\item{If $r< r_c(\phi)$, the largest component on the projection to $W$ has 
at most of the order $\log^3 L$ vertices.} 
\item{If $r> r_c(\phi)$, there is exactly one component on the projection 
with more than $T\log L$ vertices for some constant $T$, and the size of that 
component is $L \Delta + o(L)$, where
\begin{equation}
\label{Delta}
\Delta = 1 - \frac{Z((1-\beta r)\phi)}{Z(\phi)}
\end{equation}
and $\beta$ is the positive solution to
\begin{equation}
\label{beta}
\beta = 1 - \frac{Z'((1-\beta r)\phi)}{Z'(\phi)}.
\end{equation}}  
\end{enumerate}

Notice that, since $Z'((1-\beta r)\phi)$ is a convex 
function of $\beta$ on $(0,1)$, Eq.~(\ref{beta}) has a unique positive 
solution exactly when $r > r_c(\phi)$. 

We can also deduce the leading order 
of $\Delta$ as a function of $r-r_c(\phi)$: The expansion 
of the right-hand side of Eq.~(\ref{beta}) for small $\beta$ 
(all the derivatives of $Z(\phi)$ exist because $\phi<\Phi$) yields
\begin{equation}
\beta \approx \frac{2Z'(\phi)}{r^2 \phi^2 Z'''(\phi)} \left(\frac{r}
{r_c(\phi)} - 1 \right)
\end{equation}
as $r\to r_c$ from above, so that now expanding the right-hand side of 
Eq.~(\ref{Delta}) to the first order in $\beta$ we have
\begin{equation}
\label{Deltalin}
\Delta \approx r\rho \beta \approx  \frac{2 \phi Z''(\phi)^2}
{ Z(\phi) Z'''(\phi)} (r - r_c(\phi) ).
\end{equation}
Thus the growth starts linearly. Furthermore, the uniqueness of the 
solution to Eqs.~(\ref{Delta}) and (\ref{beta}) allows us to calculate 
the leading order of $\Delta$ when the critical curve is crossed 
by increasing the density $\rho$ with constant $r$. At the point 
$(\phi + \epsilon,r_c(\phi))$ with $\phi + \epsilon < \Phi$, we get
\begin{equation}
\label{Deltalin2}
\Delta \approx - \frac{2 \phi Z''(\phi)^2}
{ Z(\phi) Z'''(\phi) }r_c'(\phi) \epsilon
\end{equation}
for the size of the giant component. Note that, in case of a finite radius of 
convergence $\Phi$ of the partition function, the $\phi \to \Phi$ limits of 
the amplitudes in expressions (\ref{Deltalin}) and (\ref{Deltalin2}) depend 
crucially on the existence of the third moment of a 
$\nu_{\Phi}$-distributed random variable.  

\subsection{Two examples}
\label{SecExamples}

As a first example, let us study the case of non-interacting random walks. 
Then $g(k) =k$ and $Z(\phi) = \exp(\phi)$, so that from 
Eqs.~(\ref{defphi},\ref{setphi}) 
and from the definition of the critical density of the zero-range process, 
Eq.~(\ref{rhoc}), we get that 
$\phi = \rho$ and $\rho_c = \infty$. Also, from Eq.~(\ref{rc}), it follows 
that $r_c(\rho) = 1/\rho$, for $\rho \geq 1$, and there is no phase 
transition for $\rho < 1$ (remember that $r = 2 M_2/N \in [0,1]$). 
Of course, by $r_c(\rho)$ we mean $r_c(\phi(\rho))$ with 
$\phi(\rho)$ the solution to Eq.~(\ref{setphi}). Setting $c:= r\rho$ one 
recovers the results familiar from the model $G_{n,M=cn}$ of random graphs 
\cite{Bollobas85}: 
The phase transition occurs at $c=1$ and the size of the giant component is 
$L \Delta + o(L)$ with $\Delta$ given by the solution to 
$\Delta = 1 - \exp(-c\Delta)$. Furthermore, Eq.~(\ref{Deltalin}) is 
consistent with the known expansion $\Delta = 2\epsilon -8\epsilon^2 /3 + 
O(\epsilon^3)$ for $c= 1+\epsilon$. 

In our second example we take the jump rates to be degree independent, 
$g(k) = 1$. Now the partition function has a finite radius of convergence, 
$\Phi =1$, but still there is no condensation transition, so that 
$\rho_c =\infty$. In this case we have $r_c(\rho) = 1/(2\rho)$, for  
$\rho \geq 1/2$, and no transition for smaller densities. 
Quite surprisingly, the size of the giant component has a simple expression 
\begin{equation*}
\Delta = \frac{3}{2} - \frac{1}{2}\sqrt{1+ \frac{4}{r\rho}},
\end{equation*}  
from which, again in harmony with Eq.~(\ref{Deltalin}), we get 
$\Delta = 4\rho \epsilon /3 - 56(\rho \epsilon)^2/27 +O(\epsilon^3)$, for 
$r = 1/(2\rho) +\epsilon$. 

\subsection{General remarks on the phase diagrams} 
\label{SecGen}

The critical curve can be alternatively written in terms of the function 
$R(\phi )$ defined in Eq.~(\ref{defphi}):
\begin{equation}
\label{rcalt}
r_c(\phi) = \left( R(\phi ) - 1  +\phi \frac{R'(\phi)}{R(\phi )}\right)^{-1},
\end{equation}  
from which it follows that
\begin{equation}
\label{rcbound}
r_c(\rho) \leq \frac{1}{\rho -1}.
\end{equation}
The fact that the equation for the critical curve involves the second 
derivative of the partition function, or equivalently the first derivative 
of $R$, already tells that the phase diagram of the graph-valued 
process can be more complicated than for the zero-range process itself, 
in which the 
existence of the phase transition is simply dictated by the asymptotics of 
the ratio $Z'(\phi)/Z(\phi)$. Three different kinds of behavior can be 
extracted from the results (\ref{rc}) and (\ref{rcalt}):
\begin{enumerate}
\item
When $R(\phi) \to \infty$ as $\phi \to \Phi$, i.e $\rho_c$ 
is infinite, the critical curve decreases to zero as a function $\rho$ in 
such a way that $r_c (\rho ) > 0$ for $\rho$ arbitrarily large. 
The high density asymptotic behavior of the curve is either 
$r_c (\rho ) \sim 1/\rho$ or given by the nontrivial part in the denominator 
of (\ref{rcalt}). This point will be explored further shortly.
\item 
When the first derivative of the partition function converges as $\phi 
\to \Phi$, in that $\rho_c$ is finite, but the second derivative diverges, 
we have a condensation transition, and the critical curve approaches zero 
as $\rho$ tends to $\rho_c$ from below. Therefore, a giant component exists 
at and 
above the zero-range critical point for any non-zero $r$. Near the 
condensation transition, where $\phi$ is just below $\Phi$, the critical 
curve goes like 
\begin{equation}
\label{r_ccond}
r_c (\rho ) \approx {\frac{\rho}{\phi(\rho) R'(\phi(\rho))}}.
\end{equation} 
\item
If also the second derivative of the partition function converges, 
$\lim_{\rho \to \rho_c} r_c (\rho )$ is non-zero. In fact, 
the values of the dilution parameter $r$ being restricted to the unit 
interval, this case is further divided into two categories according to 
whether this non-zero limiting value is greater or less than $1$. For 
$\lim_{\rho \to \rho_c} r_c (\rho )>1$, no phase transition can be achieved 
by varying the dilution parameter $r$. In both cases, however, if the 
condensation occurs on a single vertex, we know that 
above $\rho_c$ there is a giant component, but the theory developed so far 
tells nothing about its existence exactly at the zero-range critical point.
\end{enumerate}

These three points illustrate the shape of the critical curve at the high 
density limit. We remark that, for small values of $\rho$, the curve is 
expected to pick up the features of the non-interacting case $g(k) =k$, 
provided that $\lim_{\rho \to \rho_c} r_c (\rho ) < 1$. 
An example of a concrete phase diagram, with 
all the features discussed here, will be given in section \ref{SecEvans}. 

We saw that the interactions in the two examples of section 
\ref{SecExamples} produced critical curves inversely 
proportional to the density. Let us now deduce, for which functions $g$ 
this proportionality holds, at least for densities high enough. 
For this purpose, set 
\begin{equation}
r_c (\phi) = {\frac{1}{a R(\phi)}},
\end{equation}
with $a > 0$, in (\ref{rcalt}). 
This yields a linearizable first order differential equation for $R$, with 
the initial condition $R'(0) = 1/g(1)$, the solution to which is 
\begin{equation}
R (\phi) = {\frac{\phi}{g(1) +(1-a)\phi}},
\end{equation} 
where $\phi \in \lbrack 0, g(1)/ (a-1) )$ for $a>1$ and 
$\phi \in \lbrack 0, \infty )$ otherwise. 
Furthermore, solving for $Z(\phi)$ 
from Eq.~(\ref{defphi}), we get 
\begin{equation}
\label{Z1perrho}
Z(\phi ) = 
\left(1 + {\frac{1-a}{g(1)}}\phi \right)^{{\frac{1}{1-a}}},
\end{equation}
which has the expansion
\begin{equation}
Z(\phi ) = 1+ {\frac{\phi}{g(1)}} + \sum_{m\geq 2} {\frac{1}{m!}} 
\left({\frac{\phi}{g(1)}} \right)^m \prod_{k=2}^m 
\lbrack (k-1)a - (k-2) \rbrack,
\end{equation}
so that 
\begin{equation}
g(k) = 
\begin{cases}
g(1), &{\textrm{for $k=1$}} \cr
 {\frac{g(1)k}{(k-1)a - (k-2)}}, &{\textrm{for $k\geq2$}}. 
\end{cases}
\end{equation}
This form of interaction function covers a range of models, which we now 
classify for different values of $a$:

For $a = 1$ or $2$ we get the examples of section \ref{SecExamples}.

For $1<a \neq 2$, the tail of $g(k)$, which determines the high density 
behavior, can be expanded in powers of $1/k$:
\begin{eqnarray}
g(k) = \sum_{n=1}^{n_{\ast}} 
\delta_{n,k} {\frac{(a-1)n}{(n-1)a - (n-2)}} 
\\ \nonumber
\ \ \ \ + \sum_{n=n_{\ast} +1}^{\infty} 
\delta_{n,k} \sum_{p=0}^{\infty}  
\left( {\frac{a - 2}{a - 1}} \right)^p {\frac{1}{n^p}}, 
\end{eqnarray}
where $n_{\ast} =\max \lbrace 0,\lbrack {\frac{a -2}{a -1}} 
\rbrack\rbrace$, and we have chosen $g(1) = a -1$ to make the radius of 
convergence equal to unity. To leading order in the large $k$ limit, $g(k)$ 
is then given by 
\begin{equation}
g(k) \sim 
\begin{cases}
1 - b(a)/k, &{\textrm{for $a \in (1,2)$}}, \cr
1 + b(a)/k, &{\textrm{for $a >2$}},
\end{cases}
\end{equation} 
where $b(a) = |(a - 2)/(a - 1)|$. 

In the first case of $1<a<2$, which interpolates between the 
cases of non-interacting edges and the degree independent rates, the range 
of $b$ as a function of $a$ is ${\mathbb{R}}_+$, 
so that the behavior 
\begin{equation}
r_c (\rho) = {\frac{1+b}{(2+b)\rho}}
\end{equation} 
is expected to hold for all $b\in (0,\infty)$ in the high density 
limit for the interaction $g(k) = 1 - b/k +O(k^{-(1+\delta)})$.  

The interesting case of $a > 2$ with decreasing interaction function 
is in \cite{Grosskinsky03} referred to as the Evans interaction due to his 
work presented in \cite{Evans00}. Now the image of 
$\lbrace a > 2 \rbrace$ under $b$ is the interval $(0,1)$---just a tiny part 
of the nontrivial parameter space $0<b<\infty$. Thus, we predict  
\begin{equation}
\label{Evanspredict}
r_c (\rho) = {\frac{1-b}{(2-b)\rho}}
\end{equation}  
for high densities in the Evans model with $0<b<1$ and something else for 
the rest of the phase diagram. This model is known to exhibit 
a condensation transition for $b>2$, in which case our critical curve 
$r_c (\rho)$ will hit zero at $\rho_c$. The region $1<b\leq 2$ 
is therefore special - the critical curve is strictly above zero, but 
falls of faster than the first inverse power in $\rho$. This seems to be 
consistent with the observation of Gro{\ss}kinsky et al. \cite{Grosskinsky03} 
that a kind of precursor of condensation transition is present on this 
interval of the parameter space. The Evans model will be the topic of the next 
section.

In the region $0<a<1$, the cases with $a\in \left( (0,1) \setminus  
\lbrace 1 - 1/(n+1): n\in {\mathbb{N}} \rbrace \right) $ can be ruled out 
immediately, since they lead to interaction functions breaking the positivity 
assumption. So we are here left with the set  
$a\in \lbrace 1 - 1/(n+1): n\in {\mathbb{N}} \rbrace$. But this time the 
assumption on bounded increments is violated: $g(k)$ would be finite for 
$k\in \lbrace 1 ,\ldots, K \rbrace$, where $K=1/(1-a)$, and infinite for 
larger values of $k$. 
Thus we don't expect to find $1/ (a\rho)$-behavior with $a<1$ in our model.   
However, it is possible to construct processes that have 
stationary distributions corresponding to the partition functions 
(\ref{Z1perrho}) by altering the dynamics only slightly. This leads to the 
notion of generalized exclusion processes \cite{Kipnis99} and, in these 
processes, the jumps to sites already occupied by $K$ particles are simply 
suppressed. Especially, the version with $K=1$ is the simple exclusion 
process \cite{Liggett1, Liggett2, DerridaPrivman}.
We would like to remark that, since for generalized exclusions the condition 
$\rho < K$ must be satisfied, the bound (\ref{rcbound}) would not be broken.

\section{Evans Interaction}
\protect\label{SecEvans}

In this section we take $g(k) = 1 + b/k$, $b\geq 0$, so that the partition 
function is given by a hypergeometric function \cite{Abramowitz72},
\begin{equation}
\label{EvansZ}
Z(\phi ) = \sum_{k=0}^{\infty} {\frac{(k!)^2 \, \Gamma(1+b)}{\Gamma(k+1+b)}} 
\cdot \frac{\phi^k}{k!}
 = F(1,1;1+b;\phi),
\end{equation}
with the radius of convergence $\Phi=1$. 
Also, from the definition (\ref{defphi}),
\begin{equation}
\label{REvans}
R(\phi) = {\frac{\phi}{1+b}}\cdot{\frac{F(2,2;2+b;\phi)}{F(1,1;1+b;\phi)}},
\end{equation}
from which the value of the critical density can be calculated 
\cite{Grosskinsky03}: 
\begin{equation}
\label{rhocEvans}
\rho_c = 
\begin{cases}
\infty, \qquad &{\textrm{for $b\leq 2$}} \cr
1/(b-2), \qquad &{\textrm{for $b > 2$}}.
\end{cases}
\end{equation}
The formula for the critical curve is, by the definition (\ref{rc}), 
\begin{equation}
\label{rcEvans}
r_c (\phi) = {\frac{2+b}{4\phi}}\cdot{\frac{F(2,2;2+b;\phi)}{F(3,3;3+b;\phi)}}.
\end{equation}

Let the size of the largest cluster in a zero-range process on $L$ 
sites be denoted by $Z_L^{\ast}$. The following facts for Evans interaction 
are taken from Ref.~\cite{Jeon00}:
\begin{enumerate}
\item{When $\rho < \rho_c$, that is $\phi < 1$, the distribution $\nu_\phi$ 
has an exponential tail, and $Z_L^{\ast}$ is at most of the order 
$\log L$ with high probability.}
\item{At the critical point, that is $\phi = 1$ and $b>2$, the distribution 
$\nu_\phi$ has a heavy tail, $\nu_1(k)\sim k^{-b}$, and $Z_L^{\ast}$ is 
exactly of the order $L^{1/(b-1)}$ with high probability. Furthermore, 
\begin{equation}
\label{critlowbound}
P( A_2 \mid A_1 ) \geq c L^{1-b}.
\end{equation}}
\item{For $\rho > \rho_c$ and $b>3$ (this is needed in the proof),
\begin{equation}
\frac{Z_L^{\ast}}{L} \longrightarrow \rho- \rho_c 
\end{equation}
in probability as $L\to \infty$, and the size of the second largest cluster 
is $o(\sqrt{L} \log^2 L)$ with high probability.}
\end{enumerate}

The results of the previous sections cover the subcritical cases 
$\rho < \rho_c$ only. As already discussed, the case number $3$ in the 
previous paragraph implies the existence of a giant component, but its size 
is not known. The only cases that the existence of a macroscopic component is 
not yet proven are the ones exactly at the zero-range criticality with $b>3$. 
This is because the second moment calculated from the distribution $\nu_1$ 
exists in this region only, and the critical curve $r_c(\phi)$ was inversely 
proportional to the second derivative of $Z(\phi)$. Therefore 
$\lim_{\phi\to 1} r_c(\phi) = 0$ if and only if $b\leq 3$. 

Next we show that, at the zero-range critical point with $b>3.51$, our 
degree sequence is with high probability such that it satisfies the conditions 
of Molloy and Reed. Evidently, the construction of the lower bound for the 
probability of the conditioning event follows the lines of the subcritical 
analysis. The result of Jeon {\it et al}. in Eq.~(\ref{critlowbound}) 
gives an estimate for $P( A_2 \mid A_1 )$ in the whole phase diagram with 
condensates, that is $b>2$, but due to the restriction to the region with 
existing second moment of the measure $\nu_1$, things are settled by the 
usual local limit theorem. Concerning the upper bounds for the deviations 
of $n(n-2) C_n/L$ around its mean $n(n-2)\nu_1^r (n)$ (see appendix 
\ref{AppA}), we see that locally the condition $b>3.01$ would suffice: 
The second moment $\sum_{n\geq 0} n^2 \nu_1^r (n) 
< M$ for some constant $M$ and, since the largest degree is $O(L^{1/(b-1)})$ 
with high probability, 
\begin{eqnarray}
& & P \left( \left\vert n(n-2) 
  \frac{C_n}{L} - n(n-2) \nu_1^r (n) \right\vert > 
  \epsilon \right) \leq 
\\ \nonumber
& & \ \ \ \ \ \ \ \ \ \ \  \exp \left( - \frac{\epsilon L}{n^2} 
f \left(\frac{\epsilon}{M} \right) \right) \longrightarrow 0
\end{eqnarray} 
for $n=O(L^{1/(b-1)})$ as $L\to \infty$. The magnitude of the error term in 
calculating
\begin{equation}
\label{sumn2D}
\sum_{n\geq 1} n(n-2) D_n/L 
\end{equation}
turns out to be crucial. Writing the estimate (\ref{Chernoffbound}) in a 
multiplicative form (this is, in fact, exactly the same bound as given in the 
appendix A of Ref.~\cite{Alon91}) and using again the inequality 
$\epsilon f(\epsilon) \geq \epsilon^2/(2+ \epsilon)$ yields 
\begin{equation}
P \left( \left\vert \frac{C_n}{L} -  \nu_1^r (n) \right\vert > 
\epsilon \nu_1^r(n) \right) \leq \exp \left( -   \nu_1^r (n)L 
\frac{ \epsilon^2}{2+ \epsilon} \right),
\end{equation}
so that we get a little sharper estimate than that given by the 
formula (\ref{Cnfluct}): 
\begin{equation}
C_n = \nu_{1}^r (n) L + o(\sqrt{\nu_{1}^r (n)} L^{1/2 +\delta}).
\end{equation} 
Since $\nu_{1}^r (n) \leq \nu_{1} (n)/r \sim n^{-b}$ for large $n$,
the error in the quantity (\ref{sumn2D}) is 
\begin{equation}
\sum_{n=1}^{O(L^{1/(b-1)})} n^2 \sqrt{\nu_{1}^r (n)} L^{- 1/2 +\delta} = 
O\left( L^{\frac{7-2b}{2(b-1)} + \delta} \right).
\end{equation}  
Thus $b > 3.51$ is enough to get rid of it. 

\subsection{Phase diagram and critical exponents}
\label{SecPhase}

\begin{figure}
\includegraphics[clip,angle=0,width=0.5\textwidth]{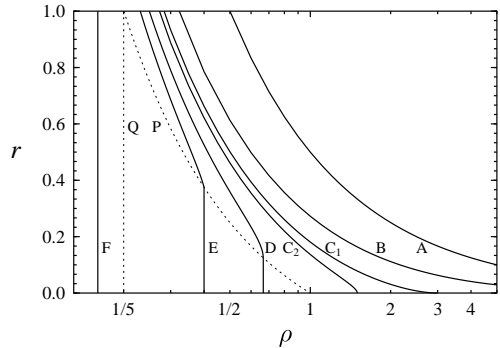}
\caption{\label{fig2} The curves A-F show phase diagrams 
for the Evans interaction with various values of $b$: From A to F, 
$b=0,3/2,7/3,8/3,7/2,9/2$ and 8. 
The dotted line P is the curve $S(\rho)$ of limiting points 
for $3< b \leq7$ and the dotted line Q marks 
the value of density $\rho$, below which the birth of the giant component is 
possible only through zero-range condensation.}   
\end{figure}

The simple approximation by independent degrees fails to give positive results 
even for the existence of the giant component for all the values of the parameter $b$. 
Now, as we start exploring the phase diagram, we present the uncertain cases 
as conjectures---they will be given some support by numerics exactly at 
criticality and analysis just below it, suggesting a cross-over between 
the subcritical and non-trivial behaviors. 
The curves A-F of Fig.~\ref{fig2} show the critical curves for a set of values 
of $b$, each representing a type of diagram different from the others. 
They will be commented one at a time as we discuss the corresponding values 
$b$.

A remark on the calculations: Since many of the results discuss the behavior 
of the system in the high density limit, in that $\phi \lesssim 1$, 
the linear transformation formula for hypergeometric functions 
($15.3.6$ of Ref.~\cite{Abramowitz72}) 
\begin{eqnarray}
\label{HGlintransf}
& & F(a_1,a_2;c;\phi) =
 \frac{\Gamma(c)\Gamma(c-a_1-a_2)}{\Gamma(c-a_1)\Gamma(c-a_2)} \times
 \\ \nonumber
& & \ \ \ \ F(a_1,a_2;a_1+a_2-c+1;1-\phi) \\ \nonumber 
& & + (1-\phi)^{c-a_1-a_2} 
\frac{\Gamma(c)\Gamma(a_1+a_2-c)}{\Gamma(a_1)\Gamma(a_2)} \times
  \\ \nonumber
& & \ \ \ \ F(c-a_1,c-a_2;c-a_1-a_2+1;1-\phi),
\end{eqnarray}
valid for $c\neq a_1+a_2+m$, $m\in {\mathbb Z}$, with expansions of the 
hypergeometric functions on the right-hand side around $\phi=1$, 
proves to be very useful. It turns out that the transformation formula is 
not valid for integer $b$ in our model, so that these situations must be dealt 
with separately by repeatedly using the Gauss' relations for contiguous 
functions given in Ref.~\cite{Abramowitz72}.

The $1/\rho$-behavior of $r_c$ 
corresponding to $\mathbf{b=0}$ is plotted as the curve A of 
Fig.~\ref{fig2}. This benchmark case has already been covered in section 
\ref{SecExamples}, so we start exploring the phase diagram from non-zero 
values of $b$. There are no condensates for $\mathbf{0<b<1}$, so that 
the results of the subcritical cases  apply. Using the above 
transformation formula and 
expanding in Eqs.~(\ref{REvans}) and (\ref{rcEvans}) one gets for high 
densities that
\begin{equation}
r_c(\rho) \approx \frac{1-b}{(2-b)\rho},
\end{equation}
exactly the same as the prediction (\ref{Evanspredict})! According to our   
calculations in section \ref{SecGen} this should be the only region, where 
the critical curve decays as a first inverse power of $\rho$. Indeed, at 
$\mathbf{b=1}$, 
\begin{equation}
\label{Zb1}
Z(\phi) = - (\log (1-\phi))/\phi,
\end{equation}
leading to a logarithmic correction 
to the first inverse power law. Furthermore, for $\mathbf{1<b<2}$, 
condensation does not occur, but we see a drastic change in the behavior of 
the critical curve,
\begin{equation}
r_c(\rho) \approx \frac{1}{2-b} \left( 
\frac{(b-1)\Gamma(2-b)\Gamma(b)}{\rho} \right)^{\frac{1}{2-b}},
\end{equation} 
in that it decays with a $b$-dependent exponent greater than $1$. 
The curve B in Fig.~\ref{fig2} with $b=3/2$ is an example from this region. 
The case $\mathbf{b=2}$ involves again logarithmic corrections. We remark 
that high density approximations for the amplitude $\Delta$ of the giant 
component could have been computed from Eq.~(\ref{Deltalin}) in the same 
manner as for the critical curves. What is important is that, by the 
subcriticality in the zero-range sense, the growth of the giant component is 
linear in the vicinity of the phase transition, i.e.~the critical exponent 
for the size of the giant component equals unity.

\begin{figure}
\includegraphics[clip,angle=0,width=0.4\textwidth]{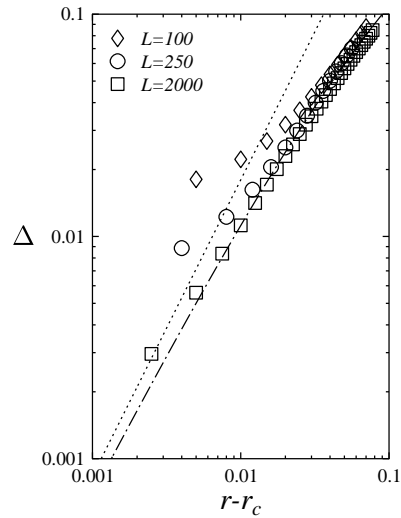}
\caption{\label{fig3} Simulation results for the size of the giant component 
at the zero-range critical point for $b=2.25$. The dash-dotted line is a 
numerical solution for the Eqs.~(\ref{Delta}) and (\ref{beta}). The dotted 
line shows the conjecture (\ref{Deltalead23}) for the leading order.}   
\end{figure}

There is a finite zero-range critical density $\rho_c = 1/(b-2)$ 
for $\mathbf{2<b<3}$, but the second derivative of the partition function 
$Z(\phi)$, i.e.~the second factorial moment of  a $\nu_{\phi}$-distributed 
variable, still diverges in the limit $\phi\to 1$. This means that $r_c$ hits 
zero at $\rho = 1/(b-2)$. Again from the expansions of the hypergeometric 
functions, we have 
\begin{equation}
r_c(\rho) \approx \frac{(b-2)^{\frac{5-2b}{b-2}}}{ \lbrack 
(b-1)\Gamma(3-b)\Gamma(b) \rbrack^{\frac{1}{b-2}}} \left( 
\frac{1}{b-2} - \rho \right)^{\frac{3-b}{b-2}}.
\end{equation}
The curves C$_1$ ($b$=7/3) and C$_2$ ($b$=8/3) in Fig.~\ref{fig2} 
with zero-range critical points $3$ and $3/2$, and high density exponents 
$2$ and $1/2$ for the critical curves, respectively, 
are examples from this region. At the exact zero-range criticality, there is 
a giant component for any non-zero $r$, but the lack of proof prevents us from 
exactly calculating its size. Certainly something interesting is happening: 
Keeping $\delta = r - r_c(\phi)$ fixed and letting $\phi\to 1$, we have  
\begin{equation}
\label{Deltalimits}
\Delta \approx  \frac{2 \phi Z''(\phi)^2}
{ Z(\phi) Z'''(\phi)} \delta \longrightarrow 
\begin{cases}
\ \ \ 0,&{\textrm{for $b\leq 4$,}} \cr
 {\frac{8(b-4)}{9(b-2)(b-3)}} \delta, &{\textrm{for $b>4$}}, 
\end{cases}
\end{equation}
so that the linear growth is expected to be taken over by some other 
behavior. To set up a conjecture, we now calculate the leading order 
approximation as if the conditions of Molloy and Reed were satisfied. 
So we wish to analyze 
\begin{equation}
\label{betaEvans}
\beta = 1 - \frac{F(2,2;2+b;(1-\beta r)\phi)}{ F(2,2;2+b;\phi) },
\end{equation}
\begin{equation} 
\label{DeltaEvans}
\Delta = 1 - \frac{F(1,1;1+b;(1-\beta r)\phi)}{ F(1,1;1+b;\phi) } 
\end{equation}
at $\phi =1$.
Expansions of the hypergeometric functions to the leading order in $\beta$ 
then yield for the size of the giant component the expression
\begin{equation}
\label{Deltalead23}
\Delta \approx \frac{1}{b-2} \left\vert \frac{\Gamma(b-2)}
{\Gamma(2-b)\Gamma(b)^2} \right\vert^{ \frac{1}{b-3} } r^{ \frac{1}{3-b} },
\end{equation}
as $r\to 0$, in that the critical exponent would be $1/(3-b)$. 
Results of Monte Carlo 
simulations, presented in Fig.~\ref{fig3} for $b=2.25$, indicate that 
the Eqs.~(\ref{betaEvans}) and (\ref{DeltaEvans}) really describe 
the size of the giant component correctly, although the convergence seems to be quite 
slow \cite{MCnote}.

To give more support to our conjecture, we remark that it can be shown 
that the non-trivial term is already present in the expansion of $\Delta$ for 
$\phi$ just below the radius of convergence. This is obtained by expanding 
the hypergeometric function in Eq.~(\ref{betaEvans}) in such a way 
that $1-\phi \ll \beta r \phi \ll 1$, in practice meaning that we are 
sufficiently far away from the critical curve measured in terms of $1-\phi$, 
and the result is the same as 
Eq.~(\ref{Deltalead23}) but with $r$ multiplied by $\phi$. 
Since the coefficient of the linear term vanishes in the limit, we then know 
that, at least for $b<2.5$, a crossover must exist at some small value of 
$r$ that vanishes in the limit $\phi\to 1$. The cases $b\geq 2.5$ would 
require a higher order analysis of the size of the giant component below the 
condensation point in order to make sure that the coefficients 
of the analytical terms with powers between $1$ and $1/(b-3)$ tend to zero 
as well.
    
The critical curve for $\mathbf{b=3}$ is of the same type as for the 
$2<b<3$ in the sense that it approaches zero as $\rho$ tends to 
$\rho_c$. The linear transformation formula being not valid for integer $b$, 
one has to deal with the exact expressions for the derivatives of the 
partition function. This implies the conjecture
\begin{equation}
\Delta \approx \exp \left(-\frac{1}{4r} \right)
\end{equation}
for the critical growth as $r\to 0$, i.e.~an infinite order transition.

Now we enter the region $\mathbf{3<b<4}$, nearly half of which allows a 
rigorous analysis. Strictly speaking, we don't even know the exact position 
of the phase transition for $b\leq 3.5$ at $\phi=1$. This is because the limit 
of the critical curve $r_c$ as $\phi\to 1$ is greater than zero by the 
existence of the second moment of a $\nu_1$-distributed variable. More 
precisely, we have $\lim_{\phi \to 1} r_c(\phi ) = (b-3)/4$, this 
holding for all $b>3$, so that, by $\rho_c = 1/(b-2)$, the end points define 
a curve
\begin{equation} 
S(\rho ) = \max \left\lbrace 0, \frac{1}{4}\left( \frac{1}{\rho} -1 \right) 
\right\rbrace 
\end{equation}
on the $(\rho, r)$-plane. Below $S(\rho)$, the transition curves are 
vertical and the giant component can be born only through condensation of 
edges. The way the curves $r_c (\rho)$ approach the zero-range 
critical point can again be evaluated,
\begin{eqnarray}
& & r_c(\rho) \approx \frac{b-3}{4} \\ \nonumber
& & + \frac{\Gamma(b)^2\Gamma(4-b)}
{16(b-3)\Gamma(b-3)} \left\lbrack \frac{(b-2)^2(b-3)}{3b-5} 
\left( \frac{1}{b-2} - \rho \right) \right\rbrack^{b-3}.
\end{eqnarray}
The curve D of Fig.~\ref{fig2} shows an example with $b=7/2$. In the figure, 
the curve P drawn with a dotted line is $S(\rho)$, the curve of limiting 
points of $r_c$. For $\mathbf{3.51<b<4}$ we can now prove that, at 
$\rho = \rho_c =1/(b-2)$ as $r\to r_c(\rho)= (b-3)/4$ from above, 
\begin{equation}
\Delta \approx \frac{1}{b-2} \left( \frac{16 \Gamma(b-2)}{(b-3)^2 \Gamma(b)^2 
\Gamma(2-b)} \right)^{\frac{1}{b-3}}  \left( r- \frac{b-3}{4} \right)^
{\frac{1}{b-3}},
\end{equation}
thus giving the exponent $1/(b-3)$, exactly the inverse of the exponent for 
the critical curve. Remember that for $2<b<3$ we conjectured $1/(3-b)$ for 
$\Delta$ but $(3-b)/(b-2)$ for $r_c(\rho)$, so that no such relation is valid 
for small values of $b$. The case $\mathbf{b=4}$ again involves 
logarithms.

We have two types of critical behavior to deal with left. The first one, 
occurring in the region $\mathbf{4<b<7}$, can be seen as a continuation 
to the preceding interval, since the endpoints of $r_c$ lie on the curve 
$S(\rho)$ with values strictly less than one because $\rho_c > 1/5$. 
The change is in the exponents: Near the zero-range criticality we have
\begin{equation}
r_c(\rho) \approx  \frac{b-3}{4} 
  + \left( \frac{9(b-3)}{4(b-4)} \!-\! 1 \right) 
\frac{(b-2)^2 (b-3)}{3b-5} \left( \frac{1}{b-2} \!-\! \rho \right),
\end{equation}
in that $r_c$ forms a cusp with the vertical line at $\rho_c$ (see the curve E 
with $b=9/2$ in Fig.~\ref{fig2}). Moreover, 
we recover the linear growth of the giant component, that we predicted already 
from the subcritical analysis in formula (\ref{Deltalimits}),
\begin{equation}
\Delta \approx \frac{8(b-4)}{9(b-2)(b-3)}  \left( r- \frac{b-3}{4} \right).
\end{equation}

The region in the phase diagram that we have not yet discussed is 
$\mathbf{b\geq7}$. But this is trivial, since $S(\rho) \geq 1$ for all values 
$\rho \leq \rho_c$, and the phase transition curves are vertical lines 
positioned at the critical density of the zero-range process (curve F in 
Fig.~\ref{fig2}). In other words, no giant component can exist without a 
condensate of edge ends on one of the vertices.

We remark that the critical exponents for the growth of the giant component at 
the condensation point are exactly the same as obtained recently by Cohen 
{\it et al}.~\cite{Cohen02} for diluted scale-free networks. This is not a 
coincidence: In their model, from a random graph with a given power-law degree 
distribution, a fraction of the vertices with all the edges joined to 
them are removed. For the vertices retained in the graph this means that their 
degree distribution is transformed exactly in the same manner as the 
distribution for the vertices in $W$ in the operation of removing 
the dangling ends from our bipartite graph. The distribution $\nu_{1}$ 
corresponding to the critical density being heavy-tailed, the similarity 
between the two problems is obvious. In the dynamical model, however, we 
have to deal with the fluctuations and, as the analysis shows, they are not 
always easily controlled.

\section{Conclusions}
\label{SecConclusions}

We studied the existence and size of the giant component on one of 
the vertex sets in a dynamical model for bipartite multigraphs using the 
results of Molloy and Reed \cite{Molloy95,Molloy98}. Motivated by the concept 
of preferential attachment, we defined the movement of the 
edge ends on one side of the bipartite graph to be a zero-range process with 
jump rates bounded away from zero, in the stationary state of which only weak 
correlations exist. A simple approximation by independent random 
variables allowed us to bound the fluctuations, and show that the degree 
sequences obtained this way are well-behaved in the sense of Molloy and Reed 
for zero-range processes with density below the critical value for 
condensation. As a benchmark, it was shown that the case of non-interacting 
edge ends produced the well-known results of the $G_{n,M}$-model of random 
graphs. Generally, we find four types of critical curves, depending on the 
asymptotics of the derivatives of the grand canonical partition function near 
the radius of convergence. A class of rate functions, for which the curves are 
of the same type as for the independent walkers, was also identified.
The last sections of the article were devoted to the Evans interaction, 
showing all the four types of phase diagrams as the parameter of the 
interaction is varied. Our simple use of the conditioning device was strong 
enough to access some nontrivial parts of the parameter space but did not 
allow us to show that the conditions of Molloy and Reed are satisfied for all 
values of the interaction parameter exactly at the 
zero-range condensation point. The critical exponents for the growth of 
the giant component were given as conjectures supported by Monte Carlo 
simulations in those cases. In the end, we discussed the connection with the 
model of diluted scale-free networks \cite{Cohen02}. This implies that our 
model with the Evans interaction and exactly at the zero-range condensation 
point belongs to the universality class of percolation on graphs with 
power-law distributed degrees. The same set of exponents appears also in 
Ref.~\cite{Goltsev03}, where a general, phenomenological Landau theory for 
phase transitions on scale-free networks is constructed. The simulation 
results suggest that the simple Landau theory is not broken by the 
fluctuations in our model.

Finally, we would like to mention a few points that would be interesting to 
analyze further. First of all, our simple 
approximations by independent variables failed to give estimates sharp 
enough to completely cover the parts of the parameter space with condensates 
for the Evans interaction. We believe that the problem is purely 
technical and could be removed by more sophisticated use of the conditioning 
device. Secondly, no results 'inside' the phase transition, i.e.\ no further 
than $o(1)$ from the critical point, exist for graphs with a general degree 
sequence. Therefore, the behavior near $r=r_c$ remains to be discovered. 
What comes to the underlying particle system, there are still open questions 
even concerning the statics of zero-range processes, especially for rate 
functions exhibiting slow decay \cite{Jeon00}. Furthermore, only stationary 
properties of the model were discussed in this article. Recently, advances 
have been made towards understanding the formation of condensates in 
zero-range processes as a function of time \cite{Godreche03,Grosskinsky03}, 
so that one is tempted to consider the temporal scalings for graph-valued 
processes as well. A grand canonical generalization of the present model 
that would allow for a variable number of external agents would also be of 
interest.

\bigskip

\begin{acknowledgments}

The authors would like to thank Pekka Kek\"al\"ainen for discussions.
This work has been supported by the Academy of Finland under 
the Center of Excellence Program (Project No.~44875). 

\end{acknowledgments}

\appendix
\section{Definitions from \cite{Molloy95,Molloy98}}
\label{AppA}

We give the definitions of Molloy and Reed in our notation. 
An asymptotic degree sequence is a sequence of integer-valued functions 
${\mathcal D}=D_0(L), D_1(L), \ldots$ such that $D_n(L)=0$, for $n\geq L$
and $\sum_{n\geq 0} D_n(L) =L$.

Given an asymptotic degree sequence ${\mathcal D}$, ${\mathcal D_L}$ is set to 
be the degree sequence ${c_1, c_2, \ldots ,c_L }$, where $c_j \geq c_{j+1}$ 
and $\vert {j: c_j = n} \vert = D_n (L)$. Let $\Omega_{\mathcal D_L}$ be 
the set of graphs with vertex set ${1,\ldots, L}$ with degree sequence 
${\mathcal D_L}$.

An asymptotic degree sequence ${\mathcal D}$ is well-behaved if:
\begin{enumerate}
\item{ ${\mathcal D}$ is feasible and smooth, i.e.\ $\Omega_{\mathcal D_L} 
\neq \emptyset $ and there are constants $\lambda_n$ 
such that $\lim_{L\to \infty} D_n(L)/L = \lambda_n$.}

\item{$n(n-2)D_n(L)/L$ tends uniformly to $n(n-2)\lambda_n$, i.e.\ for all 
$\epsilon > 0$ there exists $L_{\epsilon}$ such that for all $L>L_{\epsilon}$ 
and for all $n\geq 0$
\begin{equation*}
\vert \frac{n(n-2)D_n(L)}{L} - n(n-2)\lambda_n \vert < \epsilon
\end{equation*}
}

\item{
\begin{equation*}
{\mathcal L({\mathcal D})} = \lim_{L\to \infty} \sum_{n\geq 1} 
n(n-2)D_n(L)/L  
\end{equation*}
exists , and the sum approaches the limit uniformly, i.e.: 

\begin{enumerate}
\item{ If ${\mathcal L({\mathcal D})}$ is finite then for all $\epsilon > 0$, 
there exist $n^{\ast}$, $L_{\epsilon}$ such that for all $L>L_{\epsilon}$:
\begin{equation*}   
\vert \sum_{n=1}^{n^{\ast}} \frac{n(n-2)D_n(L)}{L} - 
{\mathcal L({\mathcal D})} \vert < \epsilon
\end{equation*}}

\item{If ${\mathcal L({\mathcal D})}$ is infinite then for all $T>0$ , 
there exist $n^{\ast}$, $L_{\epsilon}$ such that for all $L>L_{\epsilon}$:
\begin{equation*}   
\vert \sum_{n=1}^{n^{\ast}} \frac{n(n-2)D_n(L)}{L} \vert > T.
\end{equation*}}

\end{enumerate} }

\end{enumerate}

Furthermore, ${\mathcal D}$ is called sparse if $\sum nD_n(L)/L = K + o(1)$ 
for some constant $K$.

In the theorems of Molloy and Reed, ${\mathcal D}$ is assumed to be a 
well-behaved sparse asymptotic degree sequence with the property that there 
is an $\epsilon >0$ such that for all $L$ and $n> L^{1/4 -\epsilon}$, 
$D_n = 0$. In proving some statements concerning the cycles $1/4 -\epsilon$ 
has to be replaced by more restrictive $1/8 -\epsilon$. The figure $1/4$ is 
needed for the multigraph of a random configuration to be simple, and is 
therefore not required in our analysis. In fact only  $1/2 -\epsilon$, 
which is high enough for our purposes, is needed for the statements about 
the existence and size of the giant component to hold.   

Notice that our sequence ${\mathcal D}$ 
defined in section \ref{SecZRP} cannot satisfy the conditions almost surely 
(i.e.~with probability equal to $1$) but with high probability only. This 
does not matter, since all the results are given in the probabilistic setting 
anyway, and basically means a one more choice of $L_{\epsilon}$ in the proofs.

\vfill\break


\begin{thebibliography}{100}
\bibitem{whp} 
We say that an event B happens with high probability 
if $\lim_{L\to \infty} P(B)=1$.
\bibitem{MCnote}
Our main theoretical results and conjectures were tested by
performing Monte Carlo simulations, where the random number
generator Mersenne twister was utilized. For the less
interesting small values of $\rho$ the simulations were
easier than for high density, where to actually observe
the leading order critical behavior (and the exponent)
would require simulations with of the order of hundred
thousand vertices and even a larger number of particles
on $W$. Note that the parameter $M_2$ should be a
'macroscopic' number, which becomes a problem for
$r=2M_2/N$ approaching zero as in the case of Fig.~3.
\end{thebibliography}
\end{document}